\documentclass[5p]{elsarticle}
\usepackage{xcolor,soul}
\usepackage{lineno}

\usepackage{setspace}   

\usepackage[utf8]{inputenc}
\usepackage[
singlelinecheck=false 
]{caption}

\usepackage[font={small,it}]{caption}
\usepackage{amsmath}
\usepackage{upgreek}

\usepackage{wasysym}
\usepackage{amsfonts}
\usepackage{graphicx}
\usepackage{caption}
\usepackage{subcaption}
\usepackage{pgf}
\usepackage{natbib}
\usepackage{hyperref}
\usepackage{multirow}
\usepackage{booktabs}

\usepackage[colorinlistoftodos,prependcaption,textsize=tiny]{todonotes}

\begin{document}

\title{Modeling the effects of telephone nursing on healthcare utilization}

\author[label1]{Jesper Martinsson \corref{cor1}}
\ead{jesper.martinsson@ltu.se}
\address[label1]{Department of Engineering Sciences and Mathematics, Lule\r{a} University of Technology, Sweden}

\author[label2]{Silje Gustafsson}
\address[label2]{Department of Health Sciences, Lule\r{a} University of Technology, Sweden}

\cortext[cor1]{Corresponding author at: Lule\r{a} University of Technology, 971 87 Lule\r{a}, Sweden. Tel.: +46 920 491425.}

\begin{abstract} 
\noindent\emph{Background:} Telephone nursing is the first line of contact for many care-seekers and aims at optimizing the performance of the healthcare system by supporting and guiding patients to the correct level of care and reduce the amount of unscheduled visits. Good statistical models that describe the effects of telephone nursing are important in order to study its impact on healthcare resources and evaluate changes in telephone nursing procedures.\\
\noindent\emph{Objective:} To develop a valid model that captures the complex relationships between the nurse's recommendations, the patients' intended actions and the patients' health seeking behavior. Using the model to estimate the effects of telephone nursing on patient behavior, healthcare utilization, and infer potential cost savings.\\ 
\noindent\emph{Methods:} Bayesian ordinal regression modeling of data from randomly selected patients that received telephone nursing. Inference is based on Markov Chain Monte Carlo (MCMC) methods, model selection using the Watanabe-Akaike Information Criteria (WAIC), and model validation using posterior predictive checks on standard discrepancy measures. \\
\noindent\emph{Results and Conclusions:} We present a robust Bayesian ordinal regression model that predicts three-quarters of the patients' healthcare utilization after telephone nursing and we found no evidence of model deficiencies. A patient's compliance to the nurse's recommendation varies and depends on the recommended level of care, its agreement with and level of the patient's prior intention, and the availability of different care options at the time. The model reveals a risk reducing behavior among patients and the effect of the telephone nursing recommendation is 7 times higher than the effect of the patient's intended action prior to consultation if the recommendation is the highest level of care. But the effect of the nurse's recommendation is lower, or even non-existing, if the recommendation is self-care. Telephone nursing was found to have a constricting effect on healthcare utilization, however, the compliance to nurse's recommendation is closely tied to perceptions of risk, emphasizing the importance to address caller's needs of reassurance.

\end{abstract}


\maketitle

\section{Introduction}\label{sec:Introduction}
Later year’s increased patient strain on the healthcare system has increased the focus on efficiency improvements within the healthcare system. Increasing healthcare costs as well as a continuous rise in emergency department consultations for inappropriate and non-urgent conditions represent an incitement to treat illnesses at the correct level of care \citep{VSMG14}. The number of visits to the Swedish emergency departments in Stockholm are increasing by 4.5\% annually. Not only the emergency departments are experiencing an increased strain, visits to the out-of-hours clinics and primary care clinics have increased with 6.1\% and 3.1\% respectively \citep{SLL13}. This is a phenomenon that is not unique for Sweden, but can be seen in other western countries as well with up to a 40\% increase in primary care consultations during the last 20 years \citep{HV09,LEME08,VSMG14}. 

For many care-seekers, the first line of contact with healthcare services is the Swedish Healthcare Direct (SHD). The SHD is an on-call telephone nursing service similar to the NHS Direct in the UK, LINK in Canada and Health Direct in Australia. Telephone nursing  is a common work procedure for nurses in primary care, and is described by \cite{Ka09} as performing medical assessments over the telephone, while at the same time providing care with the aim of supporting, strengthening and teaching the callers and guiding care-seekers to the correct level of care. The nursing care provided is based on the caller's individual needs, and the level of care that is recommended to the callers depends on the nature of and the urgency of their symptoms. A computerized decision support tool aids the nurse in making medical assessments, and provides a shared basis for medical decisions regardless of the nurse’s experience and skills and independent of geographical location, socio-economics or other demographic factors. The system is symptom-based and designed as a checklist where questions relating to the callers symptoms is suggested, and level of urgency is indicated to the nurse \citep{Kam13}. Through a reduction of the amount of unscheduled visits, the aim of telephone nursing from an economical point of view is to improve planning and making more efficient use of healthcare resources \citep{MSM07}. But telephone nursing has been found to be highly beneficial for the caller as well in terms of receiving better information, saving time and costs as well as reducing anxiety levels \citep{GG11}. In addition to the benefits above, telephone nursing has presented many nurses with new and exciting opportunities within healthcare services and many of them enjoy their work and feel that they are offering a worthwhile service \citep{KOMM02}. 

Telephone nursing has been found to be a medically secure and cost-effective form of care, with recommendations assessed as medically adequate in 97.6\% of the cases \citep{MSM07} and with compliance rates ranging from 81.3\% \citep{MSM07} to 91\% \citep{KCH10}. Between 30-50\% of the calls result in self-care advice \citep{KCH10, MSM07, GUMA16}. Previous research has also found that care-seekers are highly satisfied with this service. However, care-seekers that receive a recommendation to practice self-care are less satisfied than those referred to medical care, indicating that their needs of care are not fully met \citep{GUMA16}.

Modeling the effects of telephone nursing is an important tool in evaluating telephone nursing's effect on healthcare utilization, as well as enabling the evaluation of possible changes in telephone nursing services.  A valid model that describes the effect of the nurse's recommendation in relation to patients' intended actions prior to consultation allows us to understand and evaluate different effects in detail. It also enables us to make predictions about reductions in healthcare costs. Earlier research \citep{GUMA16} has found that telephone nursing does not always meet care-seekers needs of care.  This underlines the importance of continuously working to improve services so that they answer to the needs of the care-seekers.

The aim of the study was to model the effects of telephone nursing. First, we wanted to find the model that best describes the effects of telephone nursing. From this model we then wanted to make inference about patient  and the effects of telephone nursing with regards to healthcare utilization and potential cost savings.

\section{Materials and methods}\label{sec:Materials_and_Methods}
In this study a model was developed and validated based on data collected in a cross-sectional study that aimed at exploring the influence of self-care advice on patient satisfaction \cite{GUMA16}. The setting of the study was the Swedish Healthcare Direct (SHD) in Northern Sweden. Northern Sweden is a large but sparsely populated area with long distances between inhabitants, hospitals and service providers. Data were collected using a questionnaire that was a further development from an existing evaluation of patient satisfaction and healthcare utilization among callers to the SHD, and details regarding the construction and content of the questionnaire has earlier been published in \cite{GUMA16}. 

With minor changes to SHD procedures, similar data could be retrieved directly from the SHD logs (in the computerized decision support tool) in combination with patients’ medical records. This means that the same model is applicable on a much larger scale, providing a national evaluation tool of the effect of SHD on healthcare utilization.

The items in the questionnaire that generated data to the model development was \emph{"What was your intended action prior to telephone nursing consultation?"}; \emph{"What action did you undertake after telephone nursing consultation?"} and \emph{"What was the nurse's recommendation?"}. Four increasing levels of care were proposed and coded as follows: 0=self-care; 1=primary care clinic; 2=out-of-hours clinic and 3=emergency department.

Eligibility criteria was all callers to the SHD in Northern Sweden during one week in the winter of 2014. A sample size calculation in \cite{GUMA16} indicated that the dispatch of 500 questionnaires would generate sufficient data. The questionnaire was dispatched to 500 randomly selected callers to the SHD, and study participants were selected using randomization lists from the SHD registers. All questionnaires were dispatched within 8 days of the call to the SHD, and were returned within 28 days of dispatch. If the respondents had made more than one call to the SHD during this week they were asked to reply regarding their last call to minimize the risk of recollection bias. Five questionnaires were returned unopened because of wrong address, and two study participants had deceased after the call. In total 225 persons returned a completed questionnaire, giving a response rate of 45.6\%. The study sample consisted of 69.3\% women and 30.7\% men, reflecting well the proportions of callers to the SHD. The mean age in the sample was 48.15 years, ranging from 17–93 years. The majority of respondents were born in Sweden (93.3\%) and cohabiting (79.9\%), and the study sample displayed a good representation of the general population in Northern Sweden (Table~\ref{tab:sweden}). The number of callers referred to the different levels of care was 75 (33.3\%) to self-care, 64 (28.4\%) to primary care; 40 (17.8\%) to out-of-hour clinic and 53 (23.6\%) to the emergency department. The complete dataset used in this study, i.e. number of individuals in each category and their corresponding explanatory variables, is retrievable from Fig.~\ref{fig:ordinal_hist}. 

\begin{table}[ht]
\caption{Characteristics of the study sample compared with the general population of Sweden.} 
\label{tab:sweden}

\footnotesize
\vspace{-0.2cm}
\begin{tabular*}{\columnwidth}{l @{\extracolsep{\fill}} lcc}
\toprule
 & Study sample & Population \\ 
 & $n=255$ (\%) & (\%) \\ 
\hline
\multicolumn{3}{l}{Occupation}\\
~~Working & 125 (56.1) & (47.4)\\
~~Student & 15 (6.7)  & (6.0) \\
~~Retired & 62 (27.8) & (22.1) \\
~~Other   & 21 (9.3)  & (11.0) \\
\multicolumn{3}{l}{Highest completed education level}\\
~~Compulsory school & 42 (19.0) & (12.0)\\
~~Upper secondary school \hspace{1cm} & 85 (38.5) & (52.0)\\
~~Tertiary education &  94 (42.5) & (36.0)\\

\bottomrule
\end{tabular*}
Source: \cite{SCB14}.
\end{table}

The outcomes in this study are patient behavior, effects on health care utilization and potential cost savings. The predicted variable is patient’s  action  undertaken  after  receiving  telephone  nursing and the explanatory variables are intended actions prior to consultation and the nurse's recommendation. 

Even though the number of explanatory variables are limited, standard ordered-logit and ordered-probit regression models \cite{Kru14,GCSR04} result in invalid descriptions of the observations in our study and fail the model validation step (described in Section~\ref{subsec:Model_validation}). If a model is not able to describe the observed data, then inference based on the model is questionable. Instead, a more flexible Bayesian ordinal regression model is proposed to describe the effects of telephone nursing on the level of care finally chosen by the caller after the consultation.

The main reasons for applying Bayesian analysis are: (i) to easily generalize standard models of ordinal data \cite{Kru14,GCSR04,GH07} (ii) to avoid non-identifiability problems in ordinal regression (e.g. separation and collinearity) by incorporating weakly informative priors proposed by \cite{GJPS08}; (iii) to provide richer inference and avoid the common problems (e.g. dependencies on the sampling and testing intentions) associated with p-values \citep{Kru13}; (iv) to make direct probability statements about the effect given the observed evidence \citep{CSAT03} (e.g. the studied effect given the observed data is inside the confidence interval with a measurable probability). 

The Bayesian ordinal regression model is a generalization of the more commonly used ordered-logit and ordered-probit regression models and adds the necessary flexibility to obtain a valid description of the observed data in this study. The considered model follows \cite{Kru14} for ordinal regression and include the following two necessary extensions to the standard logit and probit models: heterogeneous standard deviation, and additional robustness against outliers to account for possible contamination described next. These extensions are also suggested in \cite{Kru14} for robust logistic- and ordinal regression.  

Data were analyzed using the statistical packages for scientific computing with Python \citep{Oli07,MA11}. The inference is based on Markov Chain Monte Carlo methods \citep{GCSR04,Kru14,CSI00} and calculated from 100,000 samples from the posterior distribution (after burn-in) with converged chains. The confidence intervals are here based on the highest posterior density interval \cite{Kru14}.

\section{Results}\label{sec:Results}
\begin{figure*}[!ht]
\begin{center}
\includegraphics[scale=0.95]{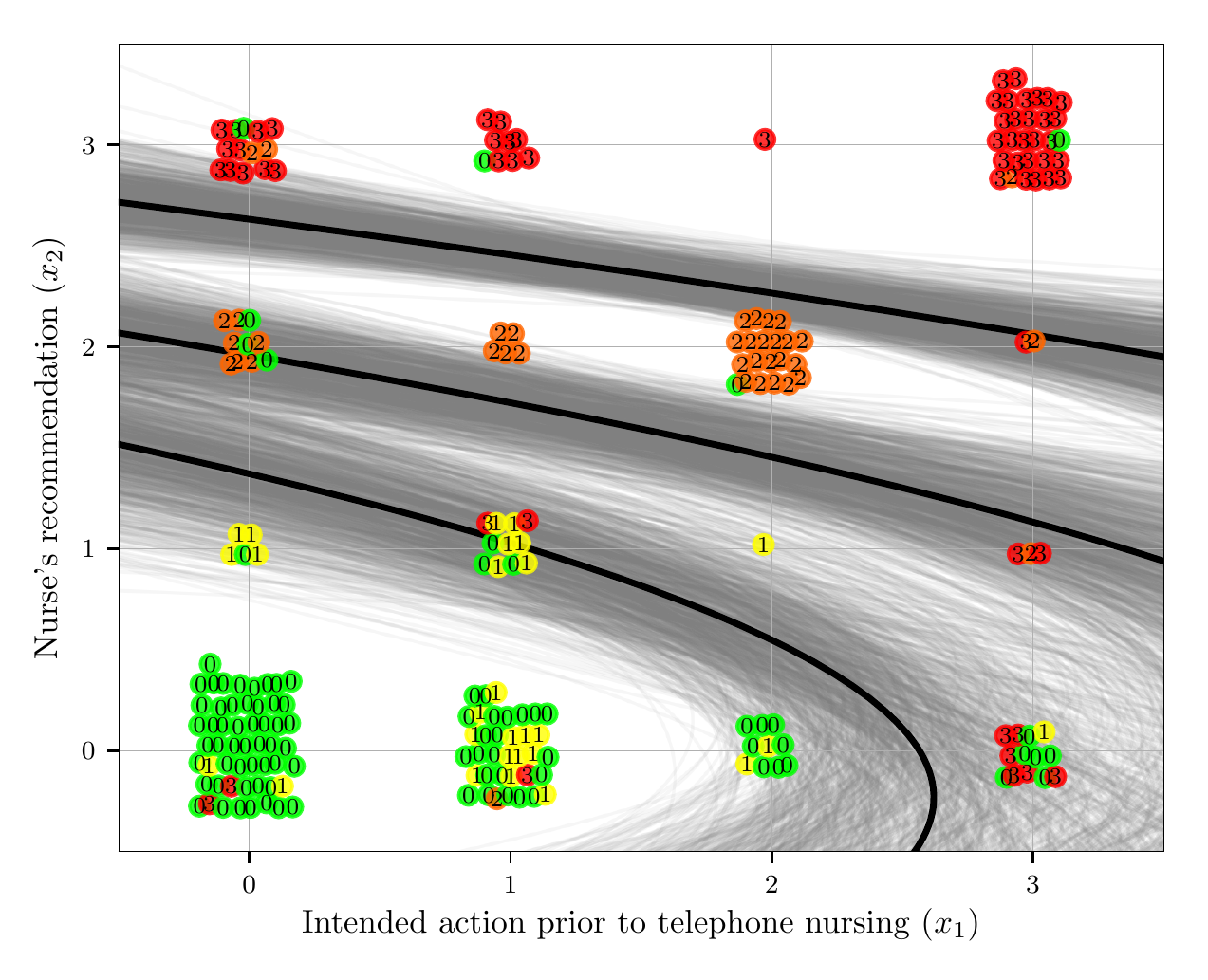}
\vspace{-0.5cm}
\caption{A scatter plot of the data and the threshold curves from the ordinal regression analysis. The color-coded numbers and the numbers on the axis represent the four increasing levels of care: 0=self-care; 1=primary care clinic; 2=out-of-hours clinic and 3=emergency department. The color-coded numbers in the scatter plot show each patient's \emph{action undertaken after receiving the nurse's recommendation} ($y_i$). The horizontal axis represents the \emph{intended action prior to consultation} ($x_{i1}$) and the vertical axis represents the \emph{nurse's recommendation} ($x_{i2}$) both categorized into the four levels of care marked by the dotted grid. For visualization purposes, the numbers are randomly scattered around their corresponding categorical integer coordinates. The thin light-gray curves depict the uncertainties involved and are credible threshold curves generated using a subset of 1000 posterior samples. }
\label{fig:ordinal}
\end{center}
\end{figure*}

The data and the results from the ordinal regression can be seen in Fig.~\ref{fig:ordinal}. The horizontal axis represents the intended action prior to consultation ($x_{1}$) and the vertical axis represents the nurse's recommendation ($x_{2}$), both categorized into the four levels of care marked by the dotted grid. The levels are defined as 0=self-care; 1=primary care clinic; 2=out-of-hours clinic and 3=emergency department. The scatter plot with color-coded numbers (0--3) shows the level of care finally chosen by each caller after receiving telephone nursing ($y$), for the same levels of care as the explanatory variables ($x_1$ and $x_2$). For visualization purposes, the numbers are randomly scattered around their corresponding categorical integer coordinates marked by the dotted grid.

\subsection{The model}\label{sec:The_model} 
The model can be summarized by the dependent variable $y_i$, which is the $i$th patient's action undertaken after receiving telephone nursing, and its two explanatory (or input) variables: $x_{i1}$ the $i$th patient's intended action prior to receiving telephone nursing; $x_{i2}$ the nurse's recommendation given to the $i$th patient. The data (i.e. $y_i$, $x_{ij}$ for $i=0,\dots,224$ patients and $j=1,2$ explanatory variables) is categorized into the four increasing levels of care defined earlier. We summarize the model in three parts below. The first part is devoted to the regression model, while the second and third parts deals with the proposed extensions to the standard ordered-logit and ordered-probit models.

\begin{itemize}
\item Part I: The mean utilized care (on the underlying metric scale) is given by a second order polynomial
\begin{linenomath*}
\begin{align}
\mu_i = \beta_0 + \beta_1 x_{i1} + \beta_2 x_{i2} + \beta_3 x_{i2}^2, \label{eq:mu}
\end{align}
\end{linenomath*}
and is the result after the model order selection procedure, described in Section~\ref{subsec:Model_selection}.
\end{itemize}

The standard ordered-logit or ordered-probit alternative assumes that the independent errors follow either a logistic or a normal distribution, respectively, \cite{Kru14,GCSR04}. However, neither distributions are flexible enough to describe the observed data in this study. Two necessary extensions to the description of the independent errors are proposed here that when combined lead to a valid description of the observations: 
\begin{itemize}
\item Part II: A single distribution for the error term is replaced with a mixture model of two distributions with different scales to account for outliers, following the recommendations in \cite{Kru14} for robust logistic- and ordinal regression. Other alternatives to increase the robustness to extreme values are available \citep[see e.g.][]{GCSR04,Kru14} but we choose the mixture model for simplicity and also for clarity as it gives a measure of the departure from the standard ordered-probit model. 
\item Part III: We also need to abandon a single common standard deviation. Instead, four different standard deviation parameters, one for each category of intended action prior to consultation, are introduced to obtain a valid description. We motivate this particular choice in Sections~\ref{subsec:Model_selection} and \ref{subsec:Inference_from_the_model}. 
\end{itemize}
In addition to providing a valid description of the observations, these extensions also contribute to interesting inferential results regarding: patient behaviour, effects of opening hours, variations in predictability, effects of perceived risks, proportion of extreme answers etc., that would otherwise not be inferred. 

We return to these tree parts in the subsequent sections below. A detailed model definition with the proposed extensions to the standard ordinal models is available in \ref{app:Model_definition}.

\subsubsection{Model validation}\label{subsec:Model_validation}
\begin{figure*}[!ht]
\begin{center}
\includegraphics[scale=0.95]{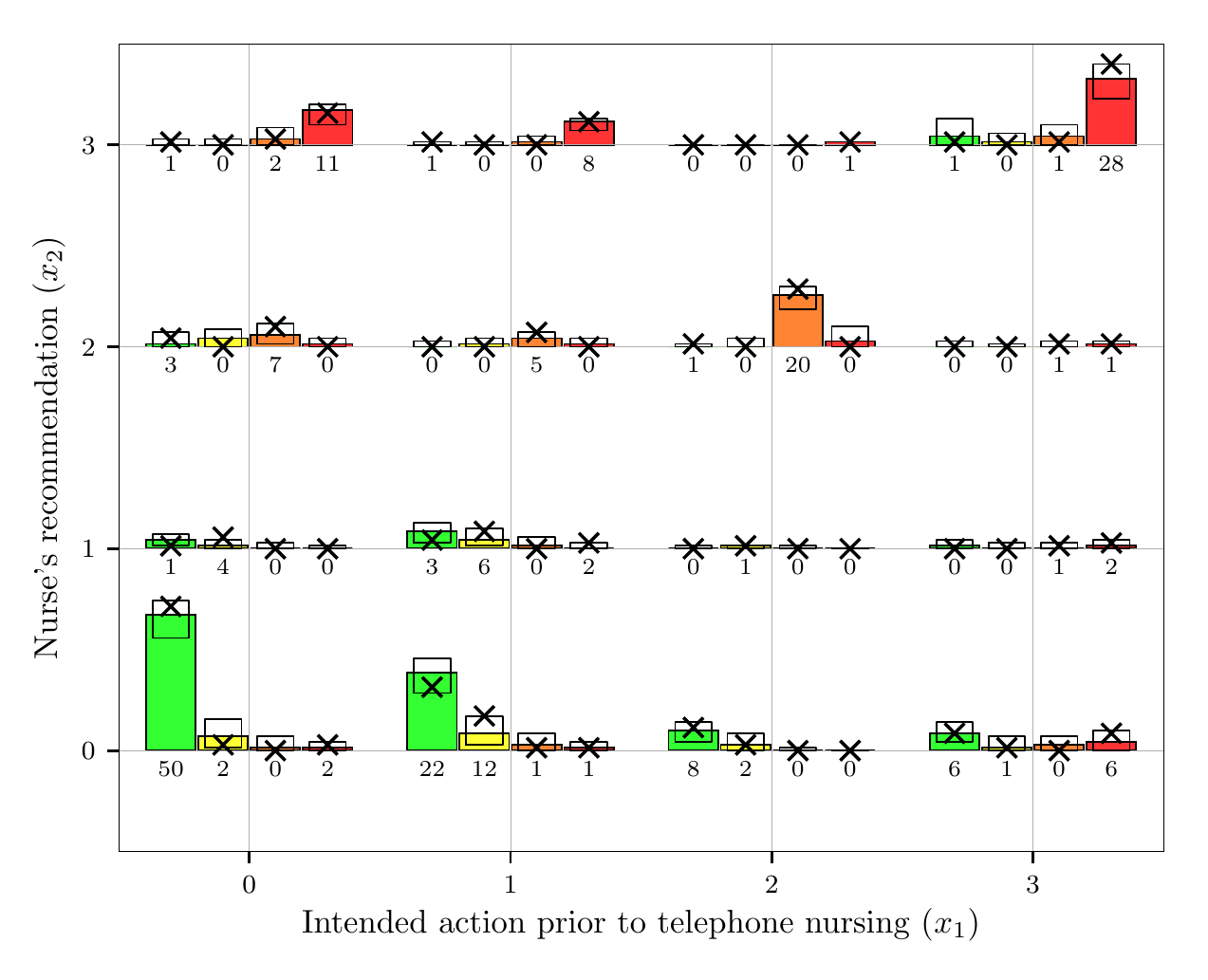}
\vspace{-0.5cm}
\caption{Histograms based on 100,000 replicated datasets from the posterior predictive distribution, seen together with a summary of the actual data (from Fig.~\ref{fig:ordinal}) represented by the black crosses. The number below each histogram bar represents the observed number of patients in each category and are the height of the black cross. The height of the histogram bar represents the mean value and the black box shows the 95\% interval of the replicated datasets from the proposed model. The 95\% interval boxes contains 63 out of the 64 crosses and 98.4 (91.7, 99.6)\% of the observations in this study.}
\label{fig:ordinal_hist}
\end{center}
\end{figure*}

Model validation is a central part in justifying the model and more importantly to justify inference based on it \citep[see e.g.][]{GCSR04,GH07}. We try to detect flaws in the model's predictive ability following the techniques recommended in \cite{GMS96,GCSR04}. In Fig.~\ref{fig:ordinal_hist}, histograms based on the model's predictions are seen together with the actual data represented by the black cross. The histogram bars represents the model's predicted level of healthcare after consultation and the black box shows the 95\% prediction interval. Based on the predictions (see \ref{app:Model_validation}), there is little evidence that would indicate significant discrepancies between the proposed model and the actual observations \citep{GCSR04}.

\subsubsection{Error rate}\label{subsec:Error_rate}
The error rate, defined as the proportion of cases for which the prediction of the final action is wrong \citep{GH07}, is 23.6\% using the proposed model. This means that the model correctly predicts 76.4\% of the patients' action undertaken after consultation, given the intended action prior to consultation and the nurse's recommendation. This can be compared to a prediction of 25\% by chance and 43.1\% using a model without the explanatory variables.

A complete probability description of all care levels was shown in Fig.~\ref{fig:ordinal_hist} and revealed no significant discrepancies between the model predictions and the actual observations.  

\subsubsection{Model selection}\label{subsec:Model_selection}
The model selection step is organized in the same three parts as was introduced earlier in Section~\ref{sec:Results}: 
\\

\noindent\textbf{Part I:} The first part is to find the appropriate combination of the explanatory variables to describe the expected level of care $\mu_i$ in (\ref{eq:mu}). All model structures up to the full quadratic form (i.e. second order polynomial including interactions) are evaluated and the model structure proposed in (\ref{eq:mu}) returns the overall lowest Watanabe-Akaike Information Criteria (WAIC) \cite{Wat10, GHV14}. The use of second order terms is also motivated by examining the clustering of the data in Fig.~\ref{fig:ordinal} with respect to the two explanatory variables, indicating non-linear behavior for large values of $x_{1}$ and low values of $x_{2}$. 
\\

\noindent\textbf{Part II:} The second part is to find the appropriate distribution to use for the error term to cope with outliers. We included outliers in the statistical model using a mixture model with a contamination parameter $\alpha$ adjusting the probability mass in the tails of the distribution, see \ref{app:Model_definition} for more details. This is the first extension to the standard ordered-logit or ordered-probit model explained previously, and is needed to pass the validation steps described in Section~\ref{subsec:Model_validation}. Including the contamination parameter also decreases both the error rate (Section~\ref{subsec:Error_rate}) and the WAIC, indicating the need of robustness against extreme answers.
\\

\noindent\textbf{Part III:} The third part is to find the appropriate combination of the explanatory variables to describe the uncertainties in the level of care ($\sigma$ in \ref{app:Model_definition}). The difference in uncertainty visible in the data in Fig.~\ref{fig:ordinal} at different values of the explanatory variables supports the use of a heterogeneous standard deviation model. The best predictive performance and lowest WAIC was given by using four separate standard deviation parameters, i.e. with a separate parameter $\sigma_{s[i]}$ for each class of intended action prior to consultation $s[i]=x_{i1}$. This choice reflects the differences in uncertainties and patients' risk reducing behaviors found in Section~\ref{subsec:Inference_from_the_model}.

\subsection{Inference from the model}\label{subsec:Inference_from_the_model}
This section starts by summarizing the model, followed by two additional subsections on healthcare utilization and cost savings.
\\

\noindent\textbf{Part I:} The regression parameters $\beta$ in (\ref{eq:mu}) describes the effect of each explanatory variable on the utilized level of care . 

The estimated effects given the observed data are summarized in Table~\ref{tab:parameters}. We see a linear positive effect ($\beta_1$) of intended action prior to consultation ($x_1$) on the utilized level of care. 
The effect of the nurse's recommendation ($x_2$) is positive and quadratic and when the nurse recommend a high level of care it is the dominating in (\ref{eq:mu}) and explain the top row of all trees at $x_2=3$ in Fig.~\ref{fig:ordinal}. From Table~\ref{tab:parameters} we conclude that:
\begin{itemize}
\item[--] If the intended action prior to consultation increases one care level, then the mean level of utilized healthcare increases 0.35 (on the underlying metric scale).
\item[--] If the nurse's recommendation increases: 
\begin{itemize}
\item[$\circ$] from self-care to primary-care, then the mean level of utilized healthcare increases 0.52 (on the underlying metric scale).
\item[$\circ$] from primary-care to out-of-hours clinic, then the mean level of utilized healthcare increases 1.24 (on the underlying metric scale).
\item[$\circ$] from out-of-hours clinic to emergency department, then the mean level of utilized healthcare increases 1.96 (on the underlying metric scale).
\end{itemize}
\end{itemize}
The relative effects on health care utilization at different combinations of $x_1$ and $x_2$ is given in Section~\ref{subsubsec:health_care_utilization}. The estimated value of $\beta_2$ is credibly zero but included as it reduces the WAIC. Also, if $\beta_3$ is significant we should include lower order terms \citep{Kru14,GH07}.

The parameter values generate the threshold curves shown in Fig.~\ref{fig:ordinal} and consequently also the outcome probabilities for each level of care given the values of the corresponding explanatory variables. The outcome probabilities are illustrated as histograms in Fig.~\ref{fig:ordinal_hist}. 
\\

\noindent\textbf{Part II:} Estimates of the contamination parameter $\alpha$ are shown in Table~\ref{tab:parameters}. As zero is outside the interval we conclude presence of extreme values in the questionnaire responses that departs from a single normal distribution. This is an additional evidence of the necessity to extend the standard ordered-logit and ordered-probit models with a mode flexible distribution of the error term. 
\\

\noindent\textbf{Part III:} The standard deviation parameters at different levels of intended action, shown in Table~\ref{tab:parameters}, depends on the caller's intended care level prior to consultation. 

The value of $\sigma_3$ is high. It seems plausible that the final action is more uncertain when the patient's intended action is the highest level of care  (3=emergency department) and the recommendation is otherwise.

The value of $\sigma_2$ in Table~\ref{tab:parameters} is low, indicating that opening hours affect both the intended action and the recommended action. As seen in the data in Fig.~\ref{fig:ordinal}, if the intended action is 2=out-of-hours clinic, it is likely that the telephone call occurred during these hours. Consequently, the intended, the recommended and the final action are more restricted with most outcomes at $x_1=x_2=2$. This prevents the spread in outcomes over all levels of care, resulting in a lower standard deviation. 
\\

\begin{table}[ht]
\caption{The mean value and 95\% confidence interval of the model parameters given the observed data.} 
\label{tab:parameters}
\centering 
\footnotesize
\vspace{-0.2cm}

\begin{tabular*}{\columnwidth}{c @{\extracolsep{\fill}} cccc}
\multicolumn{4}{ c }{The effects of the explanatory variables} \\
$\beta_0$ & $\beta_1$ & $\beta_2$ & $\beta_3$ \\ 
\midrule
-0.39          & 0.35          & 0.16          &  0.36 \\
(-0.76, -0.05) & (0.17, 0.54) & (-0.33, 0.67) & (0.15, 0.58) \\
\\
\multicolumn{4}{ c }{The uncertainties at different patient's intentions} \\
$\sigma_0$ & $\sigma_1$ & $\sigma_2$ & $\sigma_3$ \\ 
\midrule
0.71          & 0.77          & 0.28          &  2.45 \\
(0.43, 1.02) & (0.46, 1.15) & (0.18, 0.40) & (1.16, 4.19) \\
\\
\multicolumn{4}{ c }{The contamination parameter} \\
\multicolumn{4}{ c }{$\alpha$} \\
\midrule
\multicolumn{4}{ c }{0.16 } \\
\multicolumn{4}{ c }{(0.06, 0.28)} \\
\end{tabular*}
\end{table}

\subsubsection{Effects on healthcare utilization}\label{subsubsec:health_care_utilization}

The slope of the threshold curves in Fig.~\ref{fig:ordinal} reveal the \emph{relative effect} of the two explanatory variables and can be interpreted as follows:
\begin{itemize}
\item[--] If the curves are vertical, the nurse's recommendation ($x_2$) has no effect on the final action undertaken, and the patient is guided by their intended action ($x_1$) prior to consultation only.
\item[--] If the curves are horizontal, the patient's intended action prior to consultation ($x_1$) has no effect, and the patient is guided solely by the nurse's recommendation ($x_2$).
\item[--] If the curves lean with a derivative of negative one, i.e. from the upper left corners to the lower right corners in the grid, the effects of the intended action and recommended action are the same.
\end{itemize}

Overall, the threshold curves in Fig.~\ref{fig:ordinal} lean with a slope that is larger than negative one in Table~\ref{tab:derivative12} and we conclude that the nurse's recommendation is the dominating effect in the majority of the cases. The exception is when the  intended action prior to calling is the emergency department ($x_1=3$), indicating that the recommendations has little or no effect compared to the intended action. A plausible explanation is that the perceived risk to follow a self-care advice ($x_2=0$) is higher, and the patient is then guided by the intended action. 

From the reciprocal slope we conclude that if the recommendation is the emergency department ($x_2=3$) the effect is on average 7.09 times higher than the effect of the intended action prior to consultation. This can be compared to a recommendation of self-care ($x_2=0$) which gives a factor of 0.58 and credibly zero according to the 95\% interval. The relative effect increases with the nurse's recommended care-level and a plausible explanation is that disregarding a higher-level recommendation implies a higher risk compared to disregarding lower-level recommendation.

Expressions of the derivatives are available in \ref{app:Derivative}.

\begin{table}[ht]
\caption{The mean value and 95\% confidence interval of the derivatives of the threshold curves in Fig.~\ref{fig:ordinal}. Inference of the slope is given in three rows and four columns and matches the derivative of the three curves and four $x_1$ values shown in Fig.~\ref{fig:ordinal}, respectively. Inference of the reciprocal slope is seen at different care levels of the nurse's recommendation ($x_2$).} 
\label{tab:derivative12} 

\centering 
\footnotesize
\vspace{-0.2cm}
\begin{tabular*}{\columnwidth}{l @{\extracolsep{\fill}} ccccc}
\multicolumn{5}{c}{Slope $dx_2/dx_1$} \\
$k$ ~~~~ & $x_1=0$ & $x_1=1$ & $x_1=2$ & $x_1=3$ \\ \hline

\multirow{2}{*}{2} & -0.17 &  -0.18 &  -0.20 &  -0.22 \\ 
& (-0.26, -0.09) & (-0.28, -0.09) &  (-0.30, -0.09) & (-0.34, -0.09)\\ 
\midrule

\multirow{2}{*}{1} & -0.22  &  -0.25  &  -0.29  &  -$\infty$ \\
 & (-0.33, -0.12) &  (-0.38, -0.13) &  (-0.46, -0.13) & (-$\infty$, -0.18) \\
\midrule

\multirow{2}{*}{0} & -0.30  &  -0.38  &  -$\infty$  &  -$\infty$  \\ 
 & (-0.43, -0.18) &  (-0.57, -0.18) & (-$\infty$, -0.28) & (-$\infty$, -0.36) \\
 \\

\end{tabular*}

\begin{tabular*}{\columnwidth}{c @{\extracolsep{\fill}} cccc}
\multicolumn{4}{ c }{Reciprocal slope $dx_1/dx_2$} \\
$x_2=0$ & $x_2=1$ & $x_2=2$ & $x_2=3$ \\ 
\midrule

-0.58          & -2.75          & -4.92          &  -7.09 \\
 (-2.32, 1.18) & (-4.66, -1.43) & (-7.89, -2.56) & (-11.57, -3.48) \\
\end{tabular*}
\end{table}

\subsubsection{Potential cost savings \label{sec:Cost}}
The estimated cost savings in the analysis are based only on the direct costs associated with the first action undertaken by the patient after consultation and do not include additional savings as a result of improved planning of healthcare resources, reduction of unscheduled visits and guidance to the correct level of care. The results presented here are based on flat-rates from the county council where the telephone nursing consultation costs 96 SEK, a visit to the primary care clinic is approximately 2000 SEK, a visit to the out-of-hours clinic approximately 3000 SEK and a visit to the emergency department is approximately 4500 SEK.

\begin{figure}[!ht]
\begin{center}
\includegraphics[scale=1.0]{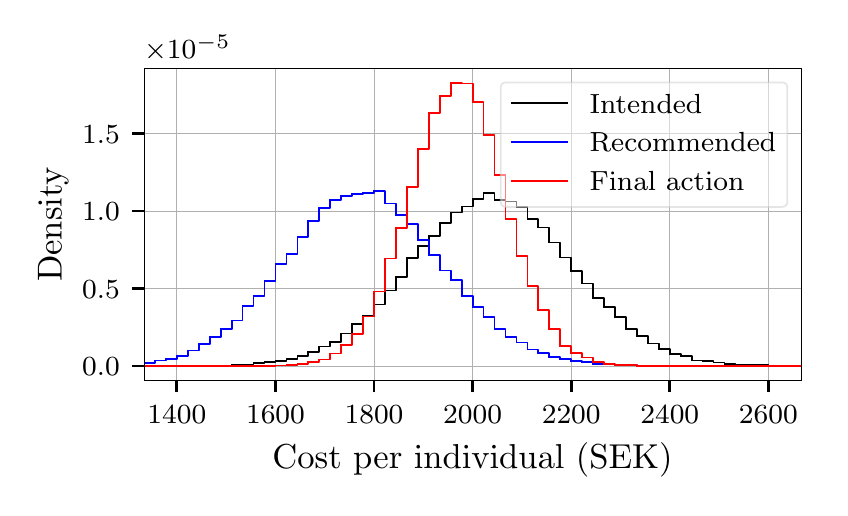}
\caption{A patient's predicted cost estimate in SEK are shown for: the intended action prior to consultation (black), the nurse's recommendation (blue) and the final action undertaken after consultation (red). The corresponding mean values and 95\% intervals are: 2039 (1731, 2365),  1779 (1462, 2081) and 1968 (1782, 2165), respectively. The actual costs associated with the observed data are 2038, 1860 and 2092, respectively.}
\label{fig:ordinal_cost}
\end{center}
\end{figure}

The predicted cost (in SEK per patient) that corresponds to the utilized healthcare after consultation can be seen in Fig.~\ref{fig:ordinal_cost} (in red). For comparison, Fig.~\ref{fig:ordinal_cost} also show both the predicted cost if all patients would follow the nurse's recommendation (in blue) and the predicted cost if the patients would do as they intended prior to consultation (in black). 

The ratio of the predicted cost associated with the final action undertaken after consultation and the intended action prior to consultation is 0.97 (0.87, 1.06) per patient. The expected cost savings associated with telephone nursing is 3.3\% compared to the intended action prior to consultation. If the patients would follow the nurse's recommendations in all cases the predicted cost ratio is 0.87 (0.71, 1.02) per patient, corresponding to an expected cost saving of 12.7\% of all patients.

\section{Discussion}\label{sec:Discussion}
Good statistical models that describe the effect of telephone nursing on healthcare utilization are essential in order to study its impact on healthcare resources and evaluate effects of future changes in telephone nursing procedures. The aim of this study was to model the effects of telephone nursing on patient behavior, healthcare utilization and infer potential cost savings. In this study we present a Bayesian ordinal regression model that captures the complex relationship between the nurse's recommendation, the patients' intended action prior to consultation and the patients' final action undertaken after receiving telephone nursing. We found no evidence to suggest any model deficiencies based on posterior predictive checks on standard discrepancy measures. 

The main finding from this study is a model that predicts 76.4\% of the patients' action undertaken after consultation. This number is based only on the most probable category predicted by the model, i.e. the model's ``best guess'', but it in addition to this it was shown that the model gives a valid probability description of all care levels after consultation.

Inference reveal that the effect of the nurse's recommendation is lower when the patient's intended action prior to consultation is the highest level of care. If the recommendation is the emergency department the effect of the recommendation is on average a factor of 7.09 times higher than the effect of the intended action prior to consultation. This can be compared to a recommendation of self-care which gives a factor of 0.58 and credibly zero according to the 95\% interval.

We found that the final action was more uncertain when the patient's intended action is the highest level of care and the nurse's recommendation is otherwise. It is even plausible that the recommendation by the nurse has no effect on the final outcome when the patient's intended action is a visit to the emergency department. The results reveal a risk reducing tendency among the callers. Consistent with several theories of health behavior, a high perceived risk of harm will according to \cite{BWC04} encourage people to take action to reduce their risk. Perceptions of risk has been found to depend on several parameters such as age, sex and health status, and that some personality types overestimate the probability of disease and/or underestimate the effects of preventive actions such as self-care \citep{EJ13}. This underlines the importance of administrating structured nursing with the specific aim of providing care-seekers with reassurance, as feeling reassured might reduce perceptions of risk. Reassurance has been found to be highly influential on patient satisfaction \citep{GUMA16}, and persons satisfied with the nurse interaction has been found to be nearly four times more likely to report engaging in self-care activities \citep{WWMJ12}. 

According to \cite{Varley201612} a clinically significant reduction in workload could be achieved if all nurses reached a proportion of patients managed by self-care of at least 65\%, assuming that referral to self-care was appropriate. A result in our study was that cost savings based only on the patient's first action undertaken after consultation is estimated to be 3.3\% but could be up to almost four times higher if compliance to the nurse's recommendations was higher. It is plausible that compliance would increase if callers were satisfied and reassured to a higher extent. The cost analysis is far from comprehensive as it neglects other significant cost reduction aspects with telephone nursing, but shows the importance of further investments to strengthen the development of the nurse-client interaction. 

In conclusion, the possibility to increase cost savings is a strong incentive to conduct further research aiming at improving persons trust in and tendency to follow the nurse's recommendations. It is likely that compliance would increase if care-seekers feel reassured and confident that following the recommendations will generate a successful outcome. This implies that the interpersonal relationship that the caller and nurse establish is highly important. To improve telephone nursing compliance, care-seekers need to feel confident that following the recommendations does not involve increased risks. More work is needed in order to further improve nurses ability to meet the needs of care-seekers.

\subsection{Study limitations}\label{subsec:Study_limitations}
The data on healthcare utilization was in this study collected through self-reporting. Social desirability bias has been found to result in higher levels of self-reported compliance to self-care advice compared to medical records \citep{OCO02}. This could imply a possible limitation as no comparison to medical records was performed. For reasons of confidentiality the researchers were not granted access to the medical records of the patients. To reduce social desirability bias, the questionnaire item regarding what action was taken was asked in a neutral manner, and this item was placed before the one asking what advice they had received from the nurse in the questionnaire.

No potential confounders were identified in our material, but it is plausible that psychological factors e.g. concern or anxiety has influenced intended actions prior to consultation. If the need of reassurance has not been adequately addressed through telephone nursing, this might have influenced the action undertaken after consultation. Psychological factors have not been analyzed due to the lack of this data in our material.  

The estimated cost savings are restricted to the direct costs associated with the first action undertaken by the patient after consultation and do not include additional savings as a result of improved planning of healthcare resources, reduction of unscheduled visits and guidance to the correct level of care. For instance, even if a patient intended to practice self-care but was recommended a visit to the emergency department, a timely guidance to the correct level-of-care may result in a reduction of the total costs in the long run. To obtain a comprehensive analysis of the savings associated with SHD, aspects such as these must be included in the analysis. 
 
The reported inference (e.g. the predictability, the specific effects of the explanatory variables, cost savings, etc.) are limited to the specific study setting and the telephone nursing procedure given by the SHD. Larger and more rigorous studies are needed to obtain evidence of effectiveness of telephone nursing nationally \cite{EKELAND20121}. However, the proposed Bayesian ordinal regression model, the model selection and validation procedure, and the interpretation of the model parameters, threshold curves, derivatives and estimands are not limited to the study setting. The reported procedures are applicable in evaluating, analyzing and understanding the effects of on-call services on healthcare utilization and patient behaviors. The number of care levels can easily be adjusted and in the simplest setting of two levels the ordinal regression model is reduced to a logistic regression \citep{Kru14}. 

With minor changes to SHD procedures, similar data could be retrieved directly from SHD logs in combination with patients’ medical records. This means that the proposed model is applicable on a much larger scale, providing not only a national evaluation tool of the effect of SHD on healthcare utilization but also more inferential evidence in combination with additional explanatory variables that may influence outcome probabilities.

\subsection{Clinical implications}\label{subsec:Clinical_Implications}
What these findings mean in a clinical setting is that telephone nursing has a constricting effect on healthcare utilization, and that guidance to the correct level of care implies a potential to save costs. However, the compliance to nurse's recommendation is closely tied to perceptions of risk, and this emphasizes the importance to address caller's needs of reassurance. A trustful relationship between the caller and the nurse carries the potential to reassure and strengthen confidence in the nurse's recommendation. But for this to happen the organization of telephone nursing needs to allow the nurse to invest time and energy to establish a trustful relationship. It also implies that nurses must acknowledge that the need for nursing care might persist even when the need for medical care is absent or resolved.

\section*{Authors' contributions}
The first author, conceived, planned, and executed the data analysis and wrote the manuscript. The second author planned and executed the data collection and contributed to writing the manuscript. Both authors have read and accepted the manuscript, and there are no conflicts of interest.

\section*{Conflicts of interest} None.


\section*{Ethical approval} 
The regional ethical review board approved this study (dnr 2010-225-31)

\section*{Summary points}

\noindent{What was already known on the topic:}
\begin{itemize}
\item There is an increase in utilization of healthcare resources that cannot be described by population growth or decreasing health. 
\item There are indications that demand and expectations of welfare services increase with increasing gross domestic product and household income.
\item On-call telephone nursing services is the first line of contact for many care-seekers and aims at optimizing the healthcare system by providing medical assessments, support and guidance to the correct level-of-care over the phone.
\end{itemize}

\noindent{What knowledge this study adds:}
\begin{itemize}
\item A valid model that describes the effect of telephone nursing in relation to patients' intended actions prior to consultation. 
\item The model is an important tool in evaluating telephone nursing's effect on healthcare utilization as well as to evaluate the effects of future changes in telephone nursing services. Inference allows us to estimate different effects in detail, understand patient behavior and patient predictability, and to estimate potential cost savings per patient. 
\item Telephone nursing has a constricting effect on healthcare utilization but the compliance to nurse's recommendation is likely to be influenced by the patients' perceptions of risk. This emphasizes the importance of meeting caller's needs of reassurance.
\end{itemize}

\section*{Acknowledgment}
The authors gratefully acknowledge all study participants for taking part in the study and Robert Lundqvist at the county council of Norrbotten for the help received with this study.

\section{References}
\bibliography{ref}

\appendix

\section{Model definition}\label{app:Model_definition}
Starting from the expression of the expected level of care given in (\ref{eq:mu}), a latent-data formulation \citep{GCSR04} of the model is given by the unobserved or latent outcome $z_i = \mu_i + \varepsilon_i$ with independent errors $\varepsilon_i\sim (1-\alpha)\mathrm{N}(0,\sigma_{s[i]})+\alpha\mathrm{N}(0,3\sigma_{s[i]})$. The errors are distributed as a mixture of two normal distributions with different scales to account for the presence of outliers. The contamination parameter $\alpha$ is estimated from the observed data (along with the other model parameters) to adjust the tail width in presence of outliers \citep{Kru14}. The observed discrete outcome 
\begin{linenomath*}
\begin{align}
y_i &=\left\{ 
\begin{array}{cl}
    0 & \mbox{if $z_i<\theta_0$}\\
    1 & \mbox{if $z_i\in [\theta_0,\theta_1]$}\\ 
    2 & \mbox{if $z_i\in [\theta_1,\theta_2]$}\\ 
    3 & \mbox{if $z_i>\theta_{2}$} 
\end{array}\right.\label{eq:y}
\end{align}
\end{linenomath*}
is categorized into the four increasing levels of care depending on the threshold values $\theta_k$ for $k=0,1,2$. 

Four different standard deviation parameters, one for each category of intended action prior to consultation, are included in the model. This is motivated by examining the deviation in the data at different $x_{1}$ levels and it also increases the model's predictive performance and decreases the Watanabe-Akaike Information Criteria (WAIC) \citep{Wat10, GHV14} compared to a homogeneous standard deviation model \citep{Kru14}. The index $s[i]=x_{i1}$ in $\sigma_{s[i]}$ returns an integer value of the category (0--3) and distinguishes the four standard deviation parameters with respect to the intended action prior to consultation ($x_1$). We return to this particular choice in Sections~\ref{subsec:Model_selection} and \ref{subsec:Inference_from_the_model}. 

The explanatory variables are mean centered and divided by the their standard deviation for computational stability \citep{Kru14}. To avoid problems with separation and collinearity we incorporate the weakly informative Cauchy prior distributions of the model parameters on the scaled explanatory variables proposed in \cite{Gel08}.
The prior distribution for the thresholds are normally distributed $\theta_k\sim N(k+1/2,2)$, where the two extreme thresholds are fixed to meaningful values on the level of care, $\theta_0=0.5$ and $\theta_2=2.5$ \citep{Kru14}.

For robust inference, the prior distribution for the contamination parameter is beta distributed $\alpha\sim\mathrm{Beta}(1,9)$, recommended in \cite{Kru14} for robust logistic- and ordinal regression. This prior distribution has a mean value of $0.1$ and gives a very low but non-zero prior probability of $\alpha>0.5$. The prior distribution for the standard deviation parameters are uniform on a very large scale.

\section{Model validation}\label{app:Model_validation}
The model generates 100,000 simulated datasets from the posterior predictive distribution and in each dataset there are 225 new individuals given the same explanatory variables as in the observed dataset. The proposed model seems to generate predicted data similar to the observed data and the 95\% confidence interval boxes in Fig.~\ref{fig:ordinal_hist} contains 63 out of the 64 crosses and 98.4 (91.7, 99.6)\% of the observations in this study. This percentage (and corresponding interval) is matches the predefined risk of 5\%, considering multiple comparisons (see e.g. \citep{GH07}).    

To further test the validity of the model we use the number of patients in each care-level as a discrepancy measure \citep{Gel13}. Based on the replicated datasets, the posterior predictive double-tailed p-vales \citep{Gel13} for the number of patients in each of the four care-levels (0=self-care, 1=primary care clinic, 2=out-of-hours clinic and 3=emergency department) are 0.412,  0.923,  0.697 and 0.185, respectively. Considering the mean value and standard deviation of the overall outcomes as discrepancy measures, return p-values of 0.195 and 0.338, respectively.

\section{Derivatives of threshold curves}\label{app:Derivative}
The curves in Fig.~\ref{fig:ordinal} are the thresholds obtained from the regression model \citep[see e.g.][]{Kru14} 
\begin{equation}
\theta_k = {\beta}_0 +{\beta}_1 x_1+{\beta}_2 x_2+{\beta}_3 x_2^2, \label{eq:threshold}
\end{equation}
for $k=0,1,2$ thresholds.

The relative effect of each explanatory variable can be investigated more precisely by examining the derivative of (\ref{eq:threshold}) with respect to each explanatory variable
\begin{align}
\frac{dx_1}{dx_2} &= \frac{-\beta_2-2\beta_3 x_2}{\beta_1},\label{eq:dthreshold1}\\
\frac{dx_2}{dx_1} &= \frac{\pm\beta_1}{\sqrt{\beta_2^2-4\beta_3(\beta_0+\beta_1 x_1-\theta_k)}},\label{eq:dthreshold2}
\end{align}
and each threshold $k=0,1,2$. The resulting derivatives are the reciprocal slope in (\ref{eq:dthreshold1}) and the slope in (\ref{eq:dthreshold2}) of the threshold curves in (\ref{eq:threshold}) and shown in Fig.~\ref{fig:ordinal}. The point and interval estimates presented in Table~\ref{tab:derivative12} are easily obtained from posterior samples of the corresponding regression and threshold parameters in (\ref{eq:dthreshold1}--\ref{eq:dthreshold1}). Note that slope (\ref{eq:dthreshold1}) is independent on $\theta_k$ and is the same for all curves at the specific $x_2$ value, this is also visible in Fig.~\ref{fig:ordinal}. The reciprocal slope, however, depends on $\theta_k$ and is different for each curve and $x_1$ value.  
\end{document}